\documentclass[fleqn,10pt]{wlscirep}
\usepackage[utf8]{inputenc}
\usepackage[T1]{fontenc}
\usepackage{float}
\usepackage{graphicx}
\usepackage{caption}
\usepackage{hyperref}
\usepackage{amsmath}
\usepackage{xcolor}
\usepackage{float}
\usepackage{placeins}
\usepackage{hypcap}
\title{PreprintToPaper dataset: connecting bioRxiv preprints with journal publications}

\author[1,*]{Fidan Badalova}
\author[2,3]{Julian Sienkiewicz}
\author[1]{Philipp Mayr}

\affil[1]{GESIS - Leibniz Institute for the Social Sciences, Cologne, 50667, Germany. \textit{\{fidan.badalova, philipp.mayr\}@gesis.org}}
\affil[2]{Warsaw University of Technology, Faculty of Physics, Warsaw, 00-662, Poland.  \textit{julian.sienkiewicz@pw.edu.pl}}
\affil[3]{Warsaw University of Technology, Centre for Credible AI, Warsaw, 00-614, Poland}

\affil[*]{Corresponding author: Fidan Badalova \textit{fidan.badalova@gesis.org}}

\begin{document}

\begin{abstract}

The PreprintToPaper dataset connects bioRxiv preprints with their corresponding journal publications, enabling large-scale analysis of the preprint-to-publication process. It comprises metadata for 145,517 preprints from two periods, 2016-2018 (pre-pandemic) and 2020-2022 (pandemic), retrieved via the bioRxiv and Crossref APIs. We selected the two periods to capture preprint–publication dynamics before and during the COVID-19 pandemic while avoiding transitional years. Each record includes bibliographic information such as titles, abstracts, authors, institutions, submission dates, licenses, and subject categories, alongside enriched publication metadata including journal names, publication dates, author lists, and further information. In addition to the main dataset, a version-history subset provides all available versions of preprints within the two selected periods, enabling analysis of how preprints evolve over time. Preprints are categorized into three groups: Published (formally linked to a journal article), Preprint Only (posted on a preprint server), and Gray Zone (potentially published in a journal but unlinked). To enhance reliability, title and author similarity scores were computed, and a human-annotated subset of 299 records was created to evaluate Gray Zone cases. The dataset supports diverse applications, including studies of scholarly communication, open science policies, bibliometric tool development, and natural language processing research on textual changes between preprints and the corresponding journal articles. The dataset is publicly available in CSV format via Zenodo\hyperlink{fn:zenodo}{\textsuperscript{*}}.

\end{abstract}

\raggedbottom
\maketitle

\begingroup
\renewcommand\thefootnote{*}
\footnotetext{\hypertarget{fn:zenodo}{}\url{https://doi.org/10.5281/zenodo.17183276}}
\endgroup

%
%
\thispagestyle{empty}

\section*{Background \& Summary}

Preprints have become an essential channel for the rapid dissemination of scientific findings, particularly in the life sciences, where timely access to results can accelerate discovery and collaboration. Their importance was intensified during the COVID-19 pandemic, when preprints served as a primary means of sharing the newest research across domains before formal peer review \cite{fraser_evolving_2021,fraser2022}.

The PreprintToPaper dataset \cite{badalova2025}, which we describe in this paper, links bioRxiv preprints with their subsequent journal publications, allowing for large-scale analysis of the preprint-to-publication process. It includes metadata on more than 145,000 preprints from two distinct periods (2016–2018, the pre-pandemic period, and 2020–2022, the COVID-19 pandemic period), with information on titles, authors, abstracts, institutions, licenses, submission and publication dates, and subject categories.  Preprints are categorized into “Published”, “Preprint Only”, and “Gray Zone” cases, based on automated linking procedures combining Digital Object Identifier (DOI) information, title similarity, and author similarity. In addition to the main dataset, two subset files are provided: (i) a human-annotated subset of Gray Zone cases is provided to support evaluation of the automated matching approach (this subset includes records with title similarity (\textit{title\_match\_score = 0.75}) and was independently checked by two annotators; further details on the annotation procedure are provided in the Methods section - Step~5: Gray Zone Verification) and (ii) a version-history subset that lists all available versions for preprints that have version one and at least one additional version. This subset is intended for detailed, version-level analyses of preprint evolution.

 The motivation for creating the PreprintToPaper dataset is to provide a comprehensive resource for studying how preprints evolve into published journal articles or remain as preprints posted only on a preprint server \cite{anderson2020,cabanac2021discovery}. While some preprints are eventually linked to formal publications, many others stay as preprints without ever appearing in journals. This dataset is the systematic effort to automatically collect and link metadata from bioRxiv \cite{sever2019} with publication records, making it possible to analyze patterns in the preprint-to-publication process at scale \cite{abdill2019tracking}. We selected the 2016–2018 (pre-pandemic) and 2020–2022 (pandemic) periods to capture preprint–publication dynamics before and during the COVID-19 pandemic while avoiding transitional years.

We identify the following potential reuse cases for the PreprintToPaper dataset.

\begin{itemize}
    \item Policy and evaluation studies: Provides evidence for assessing the role of preprints in accelerating scientific dissemination, informing open science policies, and tracking uptake during events such as the COVID-19 pandemic \cite{fraser2022,aviv-reuven2021,strcic2022}. To enable comparison, we included preprints from 2016-2018, which we designate as the pre-pandemic period.
    \item Scholarly communication research: Enables analysis of publication delays, changes in titles, abstracts, and authorship between preprints and journal articles, and differences across research fields or time periods \cite{grech2022,abdill-international2020}.
    \item Tool and method development: Serves as a reference dataset for testing algorithms in informetrics, metadata linking, and similarity matching (e.g., title and author disambiguation).
    \item Natural language processing and text analysis: Supports research on linguistic and structural changes between preprints and the corresponding journal articles, such as title reformulations, abstract rewriting, and shifts in author order or contribution statements.
\end{itemize}

In the following, we summarize related work that has used preprint data from bioRxiv.

Research on bioRxiv has shown that articles posted as preprints tend to receive more citations and online attention than comparable articles not deposited, although the causal mechanisms behind this effect remain uncertain \cite{fraser2020}. Surveys of authors indicate that the main motivations for posting preprints are to increase visibility and accelerate dissemination, while concerns about peer review and lack of awareness are the main barriers \cite{fraser2022}. Large-scale bibliometric analyses further confirm a citation and altmetric advantage for bioRxiv-linked publications, with preprints themselves increasingly cited and shared on social media \cite{fraser2020}. During the COVID-19 pandemic, preprints on bioRxiv and medRxiv played an unprecedented role, being accessed, cited, and shared at higher rates, and influencing both journalistic reporting and policy discussions \cite{fraser_evolving_2021}. An analysis of biomedical publishing during the first months of the COVID-19 pandemic reveals a sharp rise in publication volume, faster acceptance times for COVID-19 papers, and reduced international collaboration, with these shifts occurring largely at the expense of non-COVID-19 research \cite{aviv-reuven2021}.
Other studies of bioRxiv preprints have found that approximately 30\% remain as preprints posted only on a preprint server, that many are posted too late to enable substantive feedback, and that nearly half of preprints published in a journal are concentrated among four major publishers, suggesting that the platform is used more for priority setting and visibility than for pre-publication peer review \cite{anderson2020,eLife.45133}.

Below, we provide an overview of related datasets based on bioRxiv preprints and collections of COVID-19–related preprints.

The Rxivist database provides a snapshot of preprints from bioRxiv and medRxiv collected via a custom web crawler, enabling readers to sort, filter, and analyze tens of thousands of preprints based on a PostgreSQL-importable database \cite{abdill-dataset2023,abdill2019tracking}.

The covid19\_preprints dataset compiled weekly updated metadata on COVID-19–related preprints from multiple sources (Crossref, DataCite, arXiv, and RePEc), using keyword-based matching and deduplication procedures to track their distribution over time \cite{fraser-dataset2021,fraser_evolving_2021}. 

Europe PMC\cite{Levchenko2024EuropePMC} maintains a large corpus of preprints and journal articles, including bioRxiv preprints, and links them to their corresponding publications.

Compared with the three related preprint resources discussed above, the PreprintToPaper dataset offers two main advantages:
\begin{itemize}
   \item  It stores both the initial and final versions of bioRxiv preprints, along with detailed metadata (e.g., title, authors, journal, and publication date) for the corresponding  journal articles. In addition, a separate version-history file contains all available preprint versions for cases with more than one version within our study periods.
    \item It identifies articles that are not formally linked but may have been published in a journal (Gray Zone cases) by using title and author similarity, which tolerate minor title changes and variations in author order, thereby going beyond exact-match linking.
\end{itemize}

Find a detailed comparison of the four datasets in Table~\ref{tab:comparision}.

\begin{table}[H]
\centering
\caption{\label{tab:comparision} Comparison of the covid19\_preprints, Rxivist, Europe PMC, and PreprintToPaper datasets}
\resizebox{\columnwidth}{!}{%
\begin{tabular}{|p{2cm}|p{3cm}|p{3cm}|p{4.5cm}|p{4.5cm}|}
\hline
\textbf{Aspect} & \textbf{covid19\_preprints} & \textbf{Rxivist} & \textbf{Europe PMC corpus} & \textbf{PreprintToPaper (Ours)} \\
\hline
\textbf{Time span} & 2017-2021 & Nov 2013 - Feb 2023 & {2013 - 2023} & 2016-2018 (pre-pandemic) and 2020-2022 (pandemic)\\
\hline
\textbf{Coverage} & Multiple preprint servers (bioRxiv, arXiv, OSF, ESSOAR, Figshare, Zenodo, etc.) & bioRxiv and medRxiv & Multiple 32 life-science preprint servers & bioRxiv \\
\hline
\textbf{Scope} & COVID-19 related preprints only & All subjects (bioRxiv + medRxiv) & All subjects (32 servers) & All subjects (bioRxiv) \\
\hline
\textbf{Size} & 66,130 records & 223,541 records  & 761,123 records & 145,517 records \\
\hline
\textbf{Preprint information} & Source, identifier (DOI), posted date, title, abstract & Title, abstract, DOI, category, submission date, authors, institutions & Title, abstract, DOI, submission date, authors,  full text of COVID-19 and Europe PMC funder preprints in JATS XML, author affiliations, license, funding, versions, status ( Withdrawal/removal ) & Title, abstract (initial + last), DOI, category, submission date (initial + last), authors (initial + last), corresponding authors, institutions, license, version number, category, fulltext JATS XML \\
\hline
\textbf{Publication information} & None & Published journal article DOI, journal name & Published journal article DOI, journal name, publication date, citations, references & Published journal article DOI, title, authors, journal name, publication date (from Crossref) \\
\hline
\textbf{Versioning} & Only one version per preprint & Only one version per preprint & All versions (except from bioRxiv, medRxiv, SSRN) & All versions at the relevant period (bioRxiv)\\
\hline
\textbf{Publication linking} & None & Published journal article DOIs partially incomplete (crawling stopped early) & Official bioRxiv links enriched with journal metadata + an investigation of unlinked preprints where an exact match of title and first author is required. Even minor changes in title or author names are ignored; as a result, many preprints with slightly different titles are not linked to their journal papers & Official bioRxiv links enriched with Crossref + additional Gray Zone investigation where both titles and full author lists are compared using similarity scores. Preprint–paper pairs with title similarity between 75\% and 100\%  are identified and added as potential links ($\approx$19,000) \\
\hline
\end{tabular}}
\end{table}

Two further resources provide data related to preprint-publication links: PreprintMatch \cite{PreprintMatch} and the Crossref relationships dataset \cite{CrossrefRelationshipsPreprints}. In contrast to PreprintToPaper, they focus on DOI-to-DOI linkage and matching infrastructure, rather than delivering a bioRxiv-specific dataset with enriched metadata.

\section*{Methods}
\subsection*{Dataset generation workflow}
\begin{figure}[H]
\centering
\includegraphics[height=0.90\textheight,keepaspectratio]{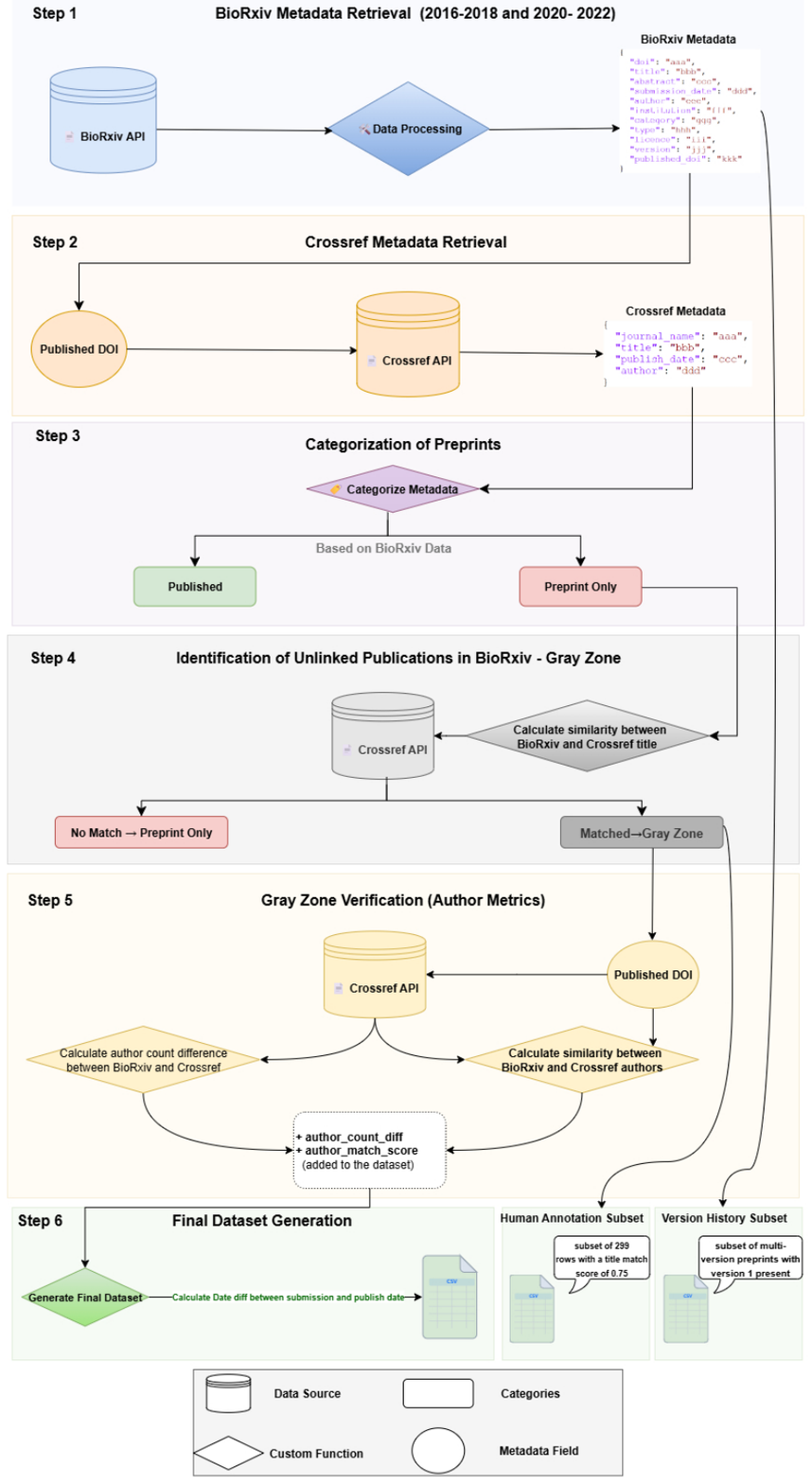}
\caption{PreprintToPaper dataset creation workflow. Final enriched dataset and two derived subsets are generated: (i) a human-annotated Gray Zone subset of 299 title-matched cases (title match score = 0.75) and (ii) a version history subset, in which multi-version preprints that retain version 1 and at least one later version within the study periods are included.}
\label{fig:workflow}
\end{figure}

Figure~\ref{fig:workflow} illustrates the data generation workflow for the PreprintToPaper dataset. The process consists of six main steps: (1) collecting metadata from the bioRxiv preprint platform; (2) retrieving additional metadata from Crossref based on the published journal article DOIs listed in the bioRxiv records; (3) categorizing preprints into published and preprint-only; (4) identifying unlinked publications in bioRxiv (Gray Zone); (5) verifying Gray Zone preprints using author-based metrics (author match score and author count difference); and (6) generating the final dataset.

\subsection*{Step 1: bioRxiv Metadata Retrieval}
\vspace{0.5em}

The first step in this study was to obtain metadata from the bioRxiv preprint platform using the official bioRxiv API (\url{https://api.biorxiv.org/details/[server]/[interval]/[cursor]/[format]}. The API has the following structure: \textit{server} indicates the source of the data (e.g., bioRxiv or MedRxiv), 
\textit{interval} indicates the desired date range, 
\textit{cursor} indicates the pagination of the results, 
and \textit{format} indicates the output format (e.g., JSON).
The following data fields were collected through the API, and the retrieved metadata fields were mapped to the following structure in the dataset  (see Table~\ref{tab:mappingBiorxiv}).

\begin{table}[H]
\centering
\caption{ \label{tab:mappingBiorxiv} Mapping of bioRxiv API fields to the dataset columns}
\label{tab:biorxiv_mapping}
\resizebox{\columnwidth}{!}{%
\begin{tabular}{|p{4.5cm}|p{6.5cm}|p{5.5cm}|}
\hline
\textbf{bioRxiv API field}  & \textbf{Dataset column} & \textbf{Description} \\
\hline
DOI & \textit{biorxiv\_doi} & Unique digital identifier of the preprint \\
Title & \textit{biorxiv\_title} & Title of the preprint \\
Authors & \textit{biorxiv\_authors} & List of all authors  \\
Corresponding author & \textit{biorxiv\_author\_corresponding} & Name of the corresponding author \\
Corresponding author institution & \textit{biorxiv\_author\_corresponding\_institution} & Institution of the corresponding author \\
Date & \textit{biorxiv\_submission\_date} & Submission date of the preprint \\
Version  & \textit{biorxiv\_version} & Version number of the preprint \\
Type & \textit{biorxiv\_type} & Document type of the preprint\\
License & \textit{biorxiv\_license} & License\\
Category & \textit{biorxiv\_category} & Subject category \\
JATS XML path & \textit{biorxiv\_jatsxml} & Link to the XML file \\
Abstract & \textit{biorxiv\_abstract} & Abstract text \\
Published & \textit{biorxiv\_published\_doi} & Publication DOI in a journal\\
\hline
\end{tabular}}
\end{table}

Data were collected for two periods: \textbf{2016--2018} (pre-pandemic period) and \textbf{2020--2022} (COVID-19 pandemic period). 
The collected metadata was then processed. Because most preprints have multiple versions, only records that met the following criteria were retained during the processing stage:  
\begin{enumerate}
    \item If the preprint is in the relevant period and both the first and last versions of the preprint are available, then both versions were kept. 
    \item If the preprint has only one version and this version is the first, then this version was also retained. 
\end{enumerate}

If only non-initial versions were available during the relevant period, the corresponding records were deleted. Additionally, a pivoting step was applied to the entire dataset to prevent \textit{DOI} duplication. 
To avoid this, each preprint was stored only once, but non-repeating fields were stored in separate columns. 
For example, the \textit{title}, \textit{authors}, \textit{corresponding author}, 
\textit{submission dates}, and \textit{abstract} of both the first and last versions were stored in new columns. 
For this purpose, the \_1st and \_last suffixes were systematically added to the column names 
(e.g., \textit{biorxiv\_title\_1st} and \textit{biorxiv\_title\_last}).

 In addition to the main table, a separate version-history subset was created that contains all available versions for preprints with a first version and at least one subsequent version within our study periods. The filtering applied to the main dataset is designed to track changes between the first and last preprint versions, as well as changes that occur upon journal publication, whereas the version-history subset supports more detailed, version-level analyses. However, because the bioRxiv API was observed to return identical abstract text across different versions of the same preprint, abstracts were not included in the version-history subset.

\subsection*{Step 2: Crossref Metadata Retrieval}
\vspace{0.5em}
For preprints that were later published as journal articles, the corresponding journal DOI is stored in the bioRxiv dataset. About two weeks after journal publication, bioRxiv automatically adds a link to the corresponding journal article (\url{https://www.biorxiv.org/about/FAQ}). Detailed metadata for these articles was retrieved via the official Crossref API (\url{https://api.crossref.org/works/[doi]}), by replacing the doi field with the DOI of the corresponding article. This query returns an extensive metadata package for the corresponding journal article. The following fields were retrieved from the Crossref database  (see Table~\ref{tab:crossref_mapping}).

\begin{table}[H]
\centering
\caption{Mapping of Crossref API fields to the dataset columns}
\label{tab:crossref_mapping}
\resizebox{\columnwidth}{!}{%
\begin{tabular}{|p{3.5cm}|p{6.5cm}|p{7.5cm}|}
\hline
\textbf{Crossref API field}  & \textbf{Dataset column} & \textbf{Description} \\
\hline
Container-title & \textit{crossref\_journal\_name} & Name of the journal in which the article was published \\
Title & \textit{crossref\_title} & Official title of the corresponding journal article \\
Author & \textit{crossref\_authors} & List of authors  \\
Published-online & \textit{crossref\_online\_publication\_date} & Date the journal article was first made available online  \\
Published-print & \textit{crossref\_issue\_online\_date} & Date associated with the volume or issue of the journal in which the article appeared  \\

\hline
\end{tabular}}
\end{table}

The official publication dates of 500 journal articles were manually compared with the online\_publication\_date and issue\_online\_date obtained from Crossref. In most cases, the online\_publication\_date matched with the actual publication date of the article.

Based on the results of manual checks, we retained the online\_publication\_date when available and used the issue\_online\_date otherwise. Both values were then merged into a single column, \textit{crossref\_publication\_date}. Additionally, a new column,  \textit{crossref\_publication\_type}, was created to indicate whether the date refers to the online or issue version.

\subsection*{Step 3: Categorization of Preprints }
\vspace{0.5em}
In the third step, the dataset was categorized based on whether the preprints had DOIs linked to journal articles on the bioRxiv preprint server. 

\begin{enumerate}
    \item If a journal article DOI was available for the preprint, the preprint was included in 
    the \textit{published} group. 

    \item If a DOI was not available, the preprint was categorized as 
    \textit{preprint only}. 
\end{enumerate}

The categorization result was stored in the column \textit{custom\_status}.

\subsection*{Step 4: Identification of unlinked publications in bioRxiv - Gray Zone}
\vspace{0.5em}

Although bioRxiv metadata is expected to include DOIs for all preprints that have been published as journal articles, verification revealed that in some cases the DOI of the corresponding journal article is missing, even when the preprint has been published in a journal. For example, the fourth version of a preprint with DOI \texttt{\href{https://www.biorxiv.org/content/10.1101/170381v4}{10.1101/170381}}
 was submitted to bioRxiv on August 6, 2020, but the same article was later published in Brain Structure and Function with DOI \texttt{\href{https://link.springer.com/article/10.1007/s00429-020-02136-0}{10.1007/s00429-020-02136-0}}
 on September 18, 2020. Nevertheless, the preprint still appears as preprint only in the bioRxiv record.

Previous studies have also documented this issue \cite{abdill2019tracking, fraser_evolving_2021, cabanac2021discovery}. For example, Abdill and 
Blekhman (2019) found that 37.5\% of 120 bioRxiv preprints were missing 
publication links. Fraser et al.\ (2021) reported that 7.6\% of 12,788 
preprints had missing publication links. Cabanac et al.\ (2021) reported that 
60.3\% of preprints marked as journal-published on medRxiv and three other servers 
lacked corresponding publication links.

To identify such cases, we compared the titles of works marked as \textit{preprint only} with the titles of the corresponding journal articles in the Crossref database. 
First, a separate request was sent to the Crossref database for each DOI, and matching titles were identified. Python's \textbf{SequenceMatcher} library was then used to assess title similarity. The algorithm determines the longest common subsequence between two texts, accounts for the total length of matching parts, and calculates a similarity coefficient ranging from 0 to 1. A coefficient close to 1 indicates a complete match, whereas a value near 0 indicates substantial differences. 

To assess our choice of similarity measure, we compared Python’s SequenceMatcher with two FuzzyWuzzy metrics on a random sample of 100 preprint and publication title pairs from our dataset, all of which were confirmed as published in a journal. In 58\% of the cases, all three measures produced identical scores, and in the remaining cases the differences were small and did not change the classification of any pair with respect to our 0.75 threshold. We therefore retained SequenceMatcher as our main measure, as it is part of the Python standard library.

\begin{figure}[H]
  \centering
  \includegraphics[width=\linewidth,trim=0 0 0 7mm,clip]{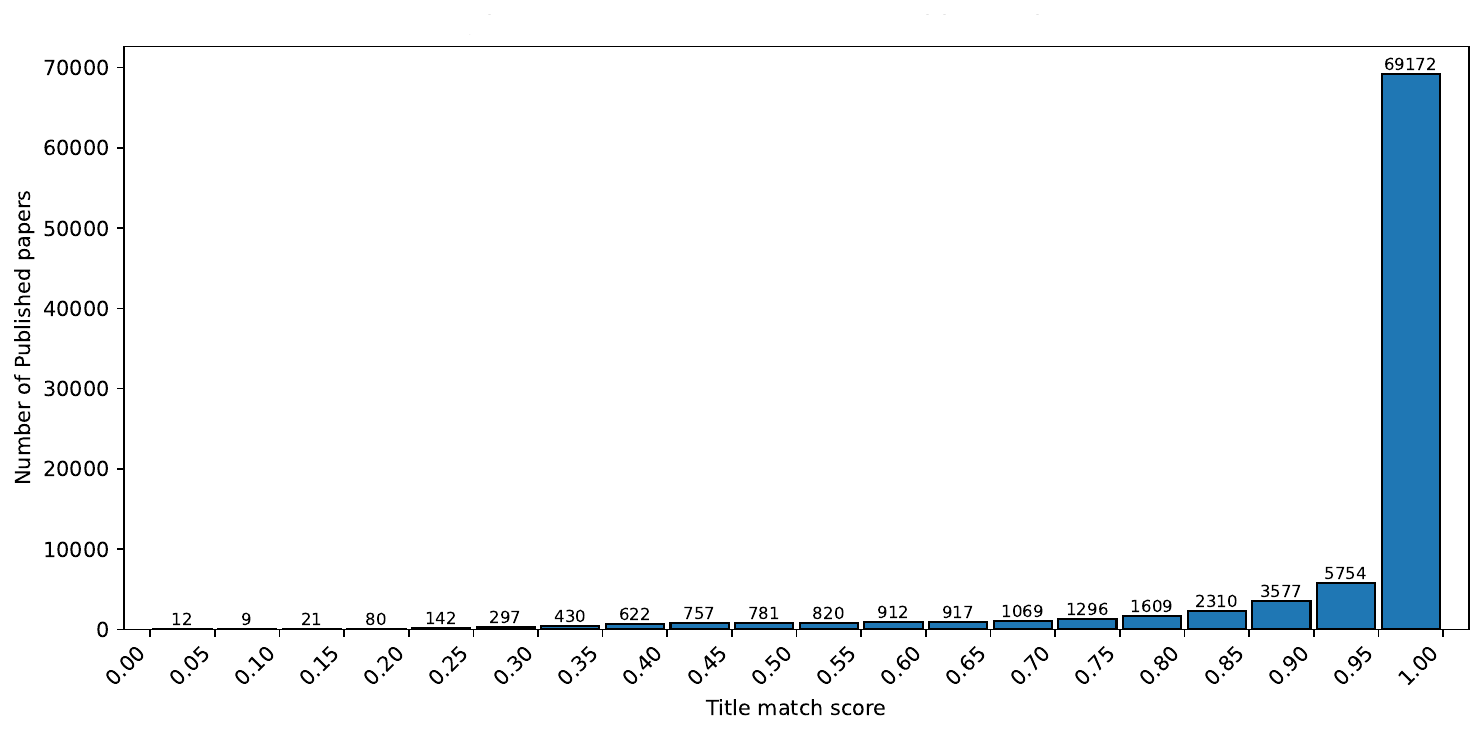}
  \caption{Distribution of journal articles by title match score.}
  \label{fig:distribution}
\end{figure}

Matches with a similarity index of $0.75$ or higher were classified into the \textit{Gray Zone} category. This threshold was chosen because, in tests comparing preprints already marked as \textit{published} in bioRxiv with their corresponding journal versions, results in this range proved most reliable in Figure~\ref{fig:distribution}.

The calculated title similarity values were stored in the \textit{title\_match\_score} column of the dataset. Additionally, for the preprints classified as \textbf{Gray Zone}, the corresponding metadata retrieved from the Crossref API including  \textit{journal name}, \textit{title}, \textit{published date}, \textit{author list}, and \textit{publication date} was stored in these fields \textit{crossref\_journal\_name}, \textit{crossref\_title}, \textit{crossref\_authors}, \textit{crossref\_publication\_date},  \textit{crossref\_publication\_type}.

\subsection*{Step 5: Gray Zone Verification (Author Metrics)}
\vspace{0.5em}
To further improve the accuracy of preprints in the 
\textit{Gray Zone} category, additional author-based metrics were applied. At this stage, the author names from the bioRxiv and Crossref metadata were compared. Two main metrics were calculated:

\begin{itemize}
    \item \textbf{Author Count Difference} -- the difference between author counts in the bioRxiv and Crossref records.
    \item \textbf{Author Match Score} -- the similarity score between the author lists.
\end{itemize}

For the second metric, Unicode normalization was first performed to minimize 
differences in name format, diacritical marks were separated and removed, and 
all letters were reduced to lowercase. Then, for each record, the author list was first divided 
into individual authors by the ``;'' separator, and each author line was tokenized 
by splitting it into words using the space character. As a result, a nested 
(i.e., list-within-a-list) structure consisting of authors was constructed for each 
line. Based on this structure, the SequenceMatcher library was applied, and 
the author's match score was calculated. This approach allowed for more accurate results in cases such as variations in the order of first and last names or when only initials were recorded.

Additionally, to validate these metrics, a human-annotated subset with a title match score of 0.75 was created, representing the lower bound of our title similarity threshold. Two annotators manually checked the corresponding bioRxiv and Crossref abstracts, and inter-annotator agreement was high.

Thus, the publication status of preprints in the Gray Zone category was determined more accurately using the author match score and author count difference metrics.

\subsection*{Step 6: Final Dataset Generation}
\vspace{0.5em}
In the final stage, several additional calculations were performed to enrich the dataset with further features. For example, the difference between the submission dates of the first and last versions of each preprint was calculated, as well as the difference between the official journal publication date and the last submission date on bioRxiv for preprints published in journal. These values were stored in the \textit{custom\_biorxivVersion\_dateDifference} and \textit{custom\_submission\&publication\_dateDiff} columns.

These operations made it possible to track the time dynamics between preprint versions and their subsequent journal publication. The finalized dataset was then saved in \texttt{CSV} format.

\section*{Data Record}

The dataset is based on data collected from the bioRxiv and Crossref databases. 
It is stored in \texttt{CSV} format and is publicly available via Zenodo \cite{badalova2025}  (\url{https://doi.org/10.5281/zenodo.17992421}). 
The dataset consists of three main files:
\begin{enumerate}
    \item The \textbf{Main dataset} includes metadata of preprints collected in 2016--2018 and 2020--2022. 
    This file contains primary fields obtained from the bioRxiv API (DOI, title, authors, corresponding institutions, abstract, version, category, license, etc.) 
    and additional fields obtained from the Crossref API (e.g., journal name, title in the journal, publication date, author list). 
    In addition, there are custom fields calculated by us: title match score, author match score, 
    author count difference, the difference between the first and last version submission dates, 
    and the difference between the last preprint submission date and the publication date. Only fields that are not self-explanatory are described in detail below:

    \begin{itemize}
  \item \textbf{\textit{custom\_status}}: Classification indicating the publication status of preprints: 
  \textit{Published} (officially published in a journal and confirmed with a DOI in bioRxiv), 
  \textit{Preprint Only} (posted on a preprint server), or \textit{Gray Zone} (no official link in bioRxiv, 
  but a possible publication identified based on title and author match).

 \item \textbf{\textit{biorxiv\_published\_doi}}: DOI of the corresponding journal article. This information 
  is collected from two sources: (1) official publication DOIs provided by bioRxiv; 
  (2) DOIs additionally identified in Gray Zone cases.

  \item \textbf{\textit{custom\_submission\&publication\_dateDiff}}: The difference in days between the last 
  preprint submission date and the publication date of the corresponding journal article.

  \item \textbf{\textit{custom\_biorxivVersion\_dateDifference}}: The difference in days between the submission 
  dates of the first and last bioRxiv versions.

  \item \textbf{\textit{biorxiv\_version\_last}}: The serial number of the last version of the preprint available 
  on bioRxiv (e.g., a value of ``5'' indicates that five versions have been published).

  \item \textbf{\textit{biorxiv\_jatsxml}}: A URL that references a JATS XML file containing the full text and 
  metadata of the preprint, if available. 

  \item \textbf{\textit{crossref\_publication\_date}}: The publication date retrieved from Crossref. This date can 
  refer either to the online publication or the issue publication.

  \item \textbf{\textit{crossref\_publication\_date\_type}}: Indicates the type of date given in the 
  \textit{crossref\_publication\_date} column: online publication or issue publication. 
  If the online publication date is available, it is preferred; otherwise, the issue 
  publication date is used.

  \item \textbf{\textit{author\_count\_diff}}: The difference between the number of authors in the preprint and 
  the number of authors in the corresponding journal article. Positive values indicate that the preprints published in journal have more authors, while negative values indicate fewer.

  \item \textbf{\textit{title\_match\_score}}: The similarity score between the preprint and the journal article titles. Values range from 0 to 1 and are calculated with the SequenceMatcher algorithm. 
  Only matches within the 0.75--1.0 range were considered potential publication candidates.

  \item \textbf{\textit{author\_match\_score}}: Similarity score between the preprint and corresponding journal article author 
  lists. Ranges from 0 to 1 and is primarily used to verify Gray Zone cases, with higher values 
  indicating stronger similarity.
\end{itemize}
    
    Table~\ref{tab:statistics} presents the distribution of preprints in the dataset by period and category 
    (Gray Zone, Preprint Only, Published) along with the total numbers.

\begin{table}[H]
\centering
\resizebox{\columnwidth}{!}{%
\begin{tabular}{|l|r|r|r|r|}
\hline
\textbf{Period} & \textbf{Gray Zone} & \textbf{Preprint Only} & \textbf{ Published} & \textbf{Total Preprints}\\
\hline
 \textbf{ 2016}       & 359       & 1~019      & 3~343 & 4~721\\
  \textbf{ 2017}       & 841       & 2~314      & 8~191 & 11~346\\
  \textbf{ 2018}      & 1~121       & 4~115     & 12~943 & 18~179\\
 \textbf{ 2020}      & 3~746       & 8~884     & 26~081 & 38~711\\
 \textbf{ 2021}      & 5~292       & 8~987     & 22~539 & 36~818\\
 \textbf{ 2022}      & 7~731      & 10~494     & 17~517 & 35~742\\\hline
 \textbf{ Total } & \textbf{ 19~090 } &  \textbf{ 35~813 } & \textbf{ 90~614 } & \textbf{ 145~517 }\\\hline
\end{tabular}}
\caption{\label{tab:statistics} Distribution of preprints in the PreprintToPaper dataset}
\end{table}

    \item The \textbf{Version History Subset} is a separate, version-level file that contains all available versions for preprints that have version one and at least one additional version within our study periods (2016–2018 and 2020–2022). Preprints with only a single version or without the first version present are not included in this file. Each row corresponds to a bioRxiv version and includes the DOI, version number, submission date, title, authors, category, license, and other core bioRxiv metadata. This subset allows detailed analyses of how preprints evolve across multiple versions.

    \item The \textbf{Human-Annotated Subset}, containing 299 records, includes only cases with a  \textit{title match score} of $0.75$. 
Preprints from both periods (2016--2018 and 2020--2022) were selected, and two annotators verified potential matches by manually checking the corresponding abstracts from bioRxiv and Crossref. To support evaluation of the Gray Zone category, this file is provided as an additional resource.
The dataset includes the following variables: \textit{year}, \textit{biorxiv\_doi}, \textit{suspected\_published\_doi}, \textit{author\_match\_score}, \textit{annotator1}, and \textit{annotator2}.
\begin{itemize}
  \item \textbf{\textit{suspected\_published\_doi}}: Suspected DOI of the journal-published article. Includes DOIs identified for Gray Zone preprints where a potential corresponding journal article was found but could not be fully confirmed. 

  \item \textbf{\textit{annotator.1 / annotator2}}: Labels provided independently by two annotators to validate Gray Zone matches. Values: \textit{True} (confirmed as correct match), \textit{False} (not a valid match), or \textit{NA} (uncertain).  
\end{itemize}
\end{enumerate}

\section*{Technical Validation}

To assess the validity of the proposed approach, all the records characterized by the $\textit{title\_match\_score} = 0.75$ (in total 299 records) have been selected and annotated by two referees; each of them had an option to mark a record as TRUE (the identified journal publication was based on the preprint), FALSE (there was no connection between the preprint and the publication) or NA (the annotator was unable to make a call). The Cohen's kappa coefficient \cite{Cohen1960} for non-NA data among annotators is equal to $\kappa=0.86$, indicating strong agreement among the annotators \cite{McHugh2012}.

\begin{figure}[H]
  \centering
  \includegraphics[width=\linewidth]{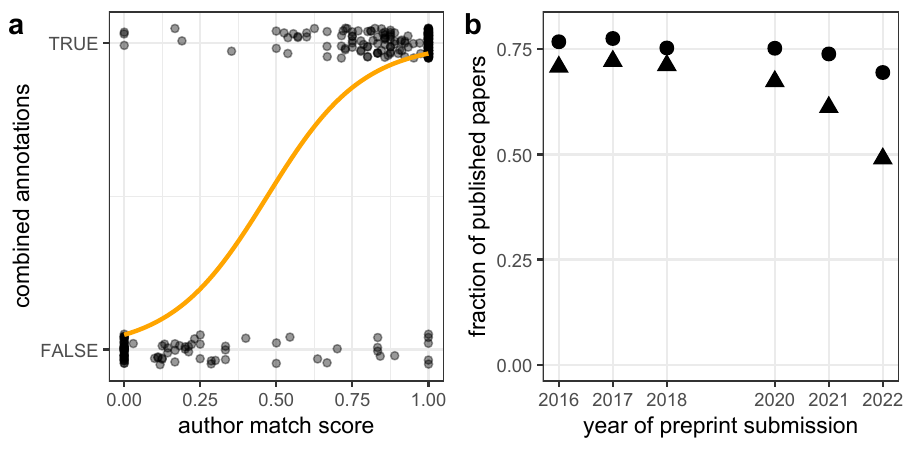}
  \caption{(a) Combined annotations versus author match score for records characterized by the title match score equal to 0.75 (see text for rationale). Symbols are data points, and the curve comes from the fit to the logistic regression model (1). Data points are artificially jittered around TRUE and FALSE values to avoid overlap. (b) The fraction of preprints that were eventually published as journal articles versus the year of preprint submission. Triangles represent data without the Gray Zone records; circles take into account Gray Zone records, with an author match score larger than 0.47 (suggested by logistic regression analysis; see text in Technical Validation) used to distinguish between preprints and journal-published articles.}
  \label{fig:lm}
\end{figure}

The annotations have been then merged by taking the conjunction of two annotations (i.e., if only if both annotators deemed the identified publication to be true, it was considered as such); if either of the annotators regarded the publication as NA it was assigned NA. The resulting distribution gives 69\% of positive recognition and 28\% of falsely assigned publications, with the remaining 3\% that cannot be resolved. These outcomes suggest that the title matching score alone might be an insufficient indicator. Figure \ref{fig:lm}a presents the combined annotation versus author match score, presenting evidence that preprint -- publication pairs with a low level of author match score are more likely to be among false positives than ones characterized by an author match score close to 1. To provide a more qualitative description of this fact and motivated by the observed relation between the combined annotation and author match score seen in Figure \ref{fig:lm}a, we used the logistic regression model\cite{Hastie2001} given by
\begin{equation}
\log\left(\frac{\pi}{1-\pi}\right)=\beta_0+\beta_1x
\label{eq:lm}
\end{equation}
where $\pi$ is interpreted as the probability of the dependent variable 
(annotation) equaling a success, $x$ reflects author match score and $\beta_0,\beta_1$ are regression coefficients. The model was fitted on the set of non-NA data, resulting in $\beta_0 = -2.99 \pm 0.46$ and $\beta_1 = 6.36 \pm 0.67$ (see the curve in Figure \ref{fig:lm}a). The location parameter $\mu = -\beta_0/\beta_1$ -- the midpoint of the curve, where $\pi$ crosses 1/2 is $\mu = 0.47 \pm 0.12$ where the standard deviation of $\mu$ has been calculated using propagation of uncertainty\cite{Ku1966}, i.e., $\sigma^2_{\mu} = |\beta_0/\beta_1|\sqrt{\sigma^2_{\beta_0} / \beta^2_0 + \sigma^2_{\beta_1} / \beta^2_1}$. To assess the validity of the logistic regression approach, we performed 10-fold cross-validation\cite{Hastie2001}, obtaining the accuracy=0.92 and Cohen's $\kappa = 0.81$.

Let us now see the ramifications of the author match score threshold $\mu$ on the possible use of the dataset. Figure \ref{fig:lm}b presents the fraction of preprints that were eventually published as journal articles $f$ versus the year of preprint submission. The first series, marked by triangles, disregards the gray zone papers, taking into account only records marked as ``published'', which leads to a spectacular decrease in $f$ by over 30\% between years 2016 and 2022. However, if we take into account that all gray zone records with author match score exceeding $\mu=0.47$ (obtained as threshold from the logistic regression analysis) can be regarded as journal-published ones, the decrease does not exceed 11\%.

\section*{Usage Notes}

PreprintToPaper dataset can be used in a variety of studies: one of the principal ideas is connected to the issue of information overload\cite{Holyst2024} in academia -- a notion that is widely understood\cite{Landhuis2016} (e.g., \textit{paper tsunami}\cite{Brainard2020}) but for which no measures have been defined yet. The very construction of our dataset (i.e., a set of preprints submitted before / during the outbreak of the COVID-19 pandemic), along with the information about the period between the preprint submission, can help find the patterns of preprint-to-paper transformations under the stress exerted by the pandemic\cite{Brainard2020}. Similarly, the characteristics of ``preprints posted on preprint server'' -- treated in a manner analogous to uncited papers\cite{Kozlowski2024,VanNoorden2017} -- can be examined using the provided dataset. These include stylometric factors: title and abstract length, text complexity (e.g., Herdan's measure\cite{Herdan1960}), readability scores (e.g., Gunning FOG index\cite{Gunning1952}) as well as text sentiment, number of authors, and other that have been shown to play role in the previous studies linking textual factors and paper popularity\cite{Didegah2013,Onodera2015,Sienkiewicz2016}. 


\section*{Data Availability}

The dataset supporting this study has been stored in Zenodo and is openly accessible at \url{https://doi.org/10.5281/zenodo.17992421}. 
Further details about the dataset contents are provided in the \textbf{Data Record} section of this manuscript.

\section*{Code Availability}

All custom scripts used in this study are publicly available in the GitHub repository: \href{https://github.com/badalovafidan/preprint_to_paper_dataset}{GitHub} (\url{https://github.com/badalovafidan/preprint_to_paper_dataset})

The principal code was developed in Python 3.9+. The main external libraries used are \textbf{pandas}, \textbf{requests}, and \textbf{numpy}. Other libraries (\textbf{unicodedata}, \textbf{re}, \textbf{difflib}, \textbf{datetime}, \textbf{time}) are included in the Python standard library. All packages required for use are listed in the \textbf{requirements.txt} file.

The code consists of scripts for different stages: downloading metadata from the bioRxiv API (\textit{download\_biorxiv.py}), selecting the first and last version and creating the pivot structure (\textit{filter\_1st\_and\_last\_version.py}, \textit{pivot\_biorxiv\_data.py}), creating additional version-history subset (\textit{create\_version\_history.py}), getting the corresponding journal article metadata from the Crossref API (\textit{download\_crossref.py}), categorizing preprints (\textit{categorize\_preprints.py}), detecting ``Gray Zone'' cases (\textit{find\_missing\_links\_crossref.py}), verifying by author match (\textit{verify\_missing\_links\_by\_author\_match.py}), standardizing dates and calculating differences (\textit{standardize\_crossref\_dates.py}, \textit{merge\_publication\_dates.py}, \textit{calculate\_date\_difference.py}).

The code for carrying out technical validation and for creating plots shown in Fig. \ref{fig:lm} is developed in \textsf{R} 4.5.1\cite{R} using the following external libraries (\textsf{R} packages): \textbf{tidyverse}\cite{tidyverse}, \textbf{cowplot}\cite{cowplot}, \textbf{psych}\cite{psych}, and \textbf{caret}\cite{caret}. The code consists of a single script \textit{technical\_validation.R}.

There are no restrictions on obtaining and using the code.

\section*{Acknowledgements}
The authors of this work were funded by the European Union under the Horizon Europe grant OMINO – Overcoming Multilevel INformation Overload (grant number 101086321, \href{http://ominoproject.eu}{http://ominoproject.eu}). Views and opinions expressed are those of the authors alone and do not necessarily reflect those of the European Union or the European Research Executive Agency. Neither the European Union nor the European Research Executive Agency can be held responsible for them. F.B. also acknowledges support from Deutsche Forschungsgemeinschaft (DFG) under grant number MA 3964/15-3, the SocioHub project. J.S. also acknowledges support from the Polish Ministry of Education and Science under the programme entitled International Co-Financed Projects and from a special grant for leaders of European projects coordinated by the Warsaw University of Technology.

\section*{Author contributions}
P.M. conceived the idea,  F.B. wrote the code, gathered the data, F.B. and J.S. annotated, validated, and analyzed the data. All authors drafted the first version of the manuscript. 

\section*{Competing interests}

The authors declare no competing interests.





\end{document}